\documentclass[12pt]{revtex4}
\pdfoutput=1
\usepackage{amssymb}
\usepackage{latexsym}
\usepackage{epsfig}
\begin{document}                             

\title{
Stability of cylindrical thin shell wormhole during  evolution of
universe from inflation to late time acceleration
 }
\author{M. R. Setare $^{1}$\footnote{rezakord@ipm.ir}, A. Sepehri $^{2}$\footnote{alireza.sepehri@uk.ac.ir}, }
\address{$^1$ Department of Science, Campus of Bijar,
University of Kurdistan, Bijar, Iran.\\ $^2$  Faculty of Physics,
Shahid Bahonar University, P.O. Box 76175, Kerman, Iran.}

\begin{abstract}

In this paper, we consider the stability of cylindrical wormholes
during evolution of universe from inflation to late time
acceleration epochs. We show that there are two types of
cylindrical wormholes. The first type is produced at the corresponding
point where k black F-strings are transited to BIon configuration. This
wormhole transfers energy from extra dimensions into our universe,
causes inflation, loses it's energy and vanishes. The second type
of cylindrical wormhole is created by a tachyonic potential and
causes a new phase of acceleration. We show that wormhole
parameters grow faster than the scale factor in this era, overtake
it at ripping time and lead to the destruction of universe at big
rip singularity.

\end{abstract}

 \maketitle
\section{Introduction}
The idea of spacetime wormhole was introduced in  1930s by
Einstein and Rosen \cite{q1} and then extended and popularized by
Morris, Thorne \cite{q2} and Visser \cite{q3}. One of important
types of these objects that may help us learn more about extra
dimensions are thin shell wormholes(TSW). These objects are
constructed by Visser from an energy- momentum at the throat
through cut-and-paste technique satisfying the proper junction
conditions. Visser proposed the first construction of thin shell
wormholes in a 3 + 1-dimensional flat Minkowski space- time
\cite{q3} and then extended  it to the Schwarzschild spacetime
\cite{q4,q5}.

Now, the main question arises what is the origin of these
wormholes in different epochs of universe? We answer to this
question in BIonic model \cite{q6,q7,qq8} which allows us to take
into account the wormhole in addition to tachyons in
brane-antibrane system. In this model, at the corresponding point
where k black fundamental strings are transited to BIon, two
universe-branes and one thin shell wormhole are born. This
wormhole is a channel for flowing energy from anti-universe-brane
to our universe and has the main role in making the inflation.
After a short period of time, wormhole loses it's energy, vanishes,
inflation ends and deceleration era begins.  With decreasing
separation between universe branes, tachyon is created and causes
 formation of the second type of unstable thin shell wormholes,
 named as tachyonic TSWs. At this stage, universe evolved from
a non-phantom deceleration era to a phantom acceleration phase and ends
up in a big rip singularity.

To construct expansion history of universe in BIon, we make use of
thin shell wormholes that have a cylindrical symmetry. Until now,
less discussions have been made on cylindrically symmetric thin
shell wormholes \cite{q8,q9,q10,q11,q12,q13,q14}. For example,
some authors presented a model  for the dynamics of non rotating
cylindrical thin-shell wormholes. They  calculated the time
evolution of the throat of special type of wormhole whose metric
has the form associated to local cosmic strings \cite{q8,q9}.
Also, they extended their calculations and built this type of
wormholes within the framework of the Brans– Dicke scalar-tensor
theory of gravity \cite{q10}. Some other authors constructed
cylindrical, traversable wormholes  by gluing two copies of a
cosmic string's manifold with a positive cosmological constant
\cite{q11}. In another scenario, authors showed that the existence
of wormholes  with cylindrical topology does not require violation
of the weak or null energy conditions near the throat \cite{q12}.
Some other investigators discussed that the wormhole and the
cosmic string geometries are locally indistinguishable, although,
their global properties are different \cite{q13}. In a recent
work, researchers investigated the stability of cylindrical thin
shell wormholes in parallel to the method  used in spherically
symmetric thin shell wormholes \cite{q14}. We will extend these
calculations to string theory and will show that these wormholes are
unstable both with the main causes of inflation and with a late time
acceleration.

 The outline of the
paper is as  follows.  In section \ref{o1}, we will consider the
evolution of cylindrical thin shell wormholes and their
annihilation during the inflation era of universe. In section
\ref{o2}, we will introduce a new type of cylindrical thin shell
wormhole, named as tachyonic TSWs. These objects were born due to
a tachyonic potential between brane-antibranes.  The last section is
devoted to summary and conclusion.

\section{ The birth and death of cylindrical wormholes during inflation}\label{o1}
In this section, we will show that cylindrical wormholes are born
at the beginning of inflation and vanish at the end of this epoch.
We obtain the location of throat in terms of BIonic parameters.

First let us  consider a general  cylindrical metric which is
given by \cite{q9}:
\begin{eqnarray}
&& ds^{2} =
B(r)(-dt^{2}+dr^{2})+C(r)d\phi^{2}+D(r)dz^{2}\label{m1}
\end{eqnarray}
in which B (r), C (r) and D(r) are as a function of r only and r=V(t)
is the location of the throat. The standard energy momentum on the
shell is:
\begin{eqnarray}
&& \rho = - (\frac{D'}{D}+\frac{C'}{C})\sqrt{\Delta}\nonumber\\
&& p_{z}=\frac{1}{\sqrt{\Delta}}[2\ddot{V}+2\frac{B'}{B}\dot{V}^{2}+\frac{B'}{B^{2}}+\frac{C'}{C}\Delta],\nonumber\\
&&
p_{\phi}=\frac{1}{\sqrt{\Delta}}[2\ddot{V}+2\frac{B'}{B}\dot{V}^{2}+\frac{B'}{B^{2}}+\frac{D'}{D}\Delta]
\label{m2}
\end{eqnarray}
where $\Delta=\frac{1}{B}+\dot{V}^{2}$. Now, we discuss the origin
of wormhole in cosmic space. In our model, coincided with the
birth of universe, wormhole is born at corresponding point where
the thermodynamics of k non-extremal black fundamental strings has
been matched to that of the BIon configuration. The supergravity
solution for k coincident non-extremal black F-strings lying along
the z direction is:
\begin{eqnarray}
&& ds^{2} = H^{-1}(-f dt^{2} + dz^{2})+ f^{-1}dr^{2} + r^{2}d\Omega_{7}^{2}\nonumber\\
&& e^{2\phi} = H^{-1},\: B_{0} = H^{-1}-1,\nonumber\\
&& H = 1 +
\frac{r_{0}^{6}sinh^{2}\alpha}{r^{6}},\:f=1-\frac{r_{0}^{6}}{r^{6}}
\label{m3}
\end{eqnarray}
  From this metric, the mass density along
the z direction can be found \cite{q7}:

\begin{eqnarray}
&& \frac{dM_{F1}}{dz} = T_{F1}k +
\frac{16(T_{F1}k\pi)^{3/2}T^{3}}{81T_{D3}}+
\frac{40T_{F1}^{2}k^{2}\pi^{3}T^{6}}{729T_{D3}^{2}}\label{m4}
\end{eqnarray}

On the other hand, for finite temperature BIon, the metric is
\cite{q6}:
\begin{eqnarray}
&& ds^{2} = -dt^{2} + dr^{2} + r^{2}(d\theta^{2} + sin^{2}\theta
d\phi^{2}) + \sum_{i=1}^{6}dx_{i}^{2}. \label{m5}
\end{eqnarray}
 Choosing the world volume coordinates of the D3-brane as
$\lbrace\sigma^{a}, a=0..3\rbrace$ and defining $\tau =
\sigma^{0},\,\sigma=\sigma^{1}$, the coordinates of BIon are given
by \cite{q6,q7}:
\begin{eqnarray}
t(\sigma^{a}) =
\tau,\,r(\sigma^{a})=\sigma,\,x_{1}(\sigma^{a})=z(\sigma),\,\theta(\sigma^{a})=\sigma^{2},\,\phi(\sigma^{a})=\sigma^{3}
\label{m6}
\end{eqnarray}
and the remaining coordinates $x_{i=2,..6}$ are constant. The
embedding function $z(\sigma)$  describes the bending of the
brane. Let z be a transverse coordinate to the branes and $\sigma$
be the radius on the world-volume.  The induced metric on the
brane is then:
\begin{eqnarray}
\gamma_{ab}d\sigma^{a}d\sigma^{b} = -d\tau^{2} + (1 +
z'(\sigma)^{2})d\sigma^{2} + \sigma^{2}(d\theta^{2} +
sin^{2}\theta d\phi^{2}) \label{m7}
\end{eqnarray}
so that the spatial volume element is $dV_{3}=\sqrt{1 +
z'(\sigma)^{2}}\sigma^{2}d\Omega_{2}$. We impose the two boundary
conditions $z(\sigma)\rightarrow 0$ for $\sigma\rightarrow
\infty$ and $z'(\sigma)\rightarrow -\infty$ for $\sigma\rightarrow
\sigma_{0}$, where $\sigma_{0}$ is the minimal two-sphere radius
of the configuration. For this BIon, the mass density along the z
direction can be obtained \cite{q7}:
\begin{eqnarray}
&& \frac{dM_{BIon}}{dz} = T_{F1}k + \frac{3\pi T_{F1}^{2}k^{2}
T^{4}}{32T_{D3}^{2}\sigma_{0}^{2}}+
 \frac{7\pi^{2} T_{F1}^{3}k^{3} T^{8}}{512T_{D3}^{4}\sigma_{0}^{4}}\label{m8}
\end{eqnarray}
Comparing the mass densities for BIon  to the mass density for the
F-strings, we see that the thermal BIon configuration behaves like
k F-strings at $\sigma = \sigma_{0}$. At this corresponding point,
$\sigma_{0}$ should have the following dependence on the
temperature:
\begin{eqnarray}
&& \sigma_{0} =
(\frac{\sqrt{kT_{F1}}}{T_{D3}})^{1/2}\sqrt{T}[C_{0} +
C_{1}\frac{\sqrt{kT_{F1}}}{T_{D3}}T^{3}]\label{m9}
\end{eqnarray}
where $T_{F1} = 4k\pi^{2}T_{D3}g_{s}l_{s}^{2}$, $C_{0}$, $C_{1}$,
$F_{0}$, $F_{1}$ and $F_{2}$ are numerical coefficients which can
be determined by requiring that $T^{3}$ and $T^{6}$ terms in
Eqs. (\ref{m4}) and (\ref{m8}) are consistent with each other. At this point, two universes
and one wormhole have been born.  The metric of these FRW
universes is:
\begin{eqnarray}
&& ds^{2}_{Uni1} = ds^{2}_{Uni2} = -dt^{2} + a(t)^{2}(dx^{2} +
dy^{2} + dz^{2}), \label{m10}
\end{eqnarray}
We assume that universes are located on D3-branes and don't have
any contribution in mass density along z direction. For this
reason, we write:
\begin{eqnarray}
&& p_{z}= \frac{dM_{F1}}{dz}\rightarrow \nonumber\\
&&\frac{1}{\sqrt{\Delta}}[2\ddot{V}+2\frac{B'}{B}\dot{V}^{2}+\frac{B'}{B^{2}}+\frac{C'}{C}\Delta]
=\nonumber\\
&& T_{F1}k + \frac{16(T_{F1}k\pi)^{3/2}T^{3}}{81T_{D3}}+
\frac{40T_{F1}^{2}k^{2}\pi^{3}T^{6}}{729T_{D3}^{2}}\label{m11}
\end{eqnarray}
Assuming $B'(V_{0})=C'(V_{0})=0$, we can solve the above equation:
\begin{eqnarray}
&& V(t)=exp(\int P_{0}dt)\nonumber\\
&& P_{_{0}}=T_{F1}k + \frac{16(T_{F1}k\pi)^{3/2}T^{3}}{81T_{D3}}+
\frac{40T_{F1}^{2}k^{2}\pi^{3}T^{6}}{729T_{D3}^{2}}\label{m12}
\end{eqnarray}
Using this equation and equation (\ref{m9}) and assuming that
throat of a stringy wormhole is equal to throat of a cylindrical
wormhole $V_{0}=\sigma_{0}$, we can derive the universe
temperature in terms of time:
\begin{eqnarray}
&& V=\sigma_{0}\rightarrow \nonumber\\
&& exp(\int
P_{0}dt)=(\frac{\sqrt{kT_{F1}}}{T_{D3}})^{1/2}\sqrt{T}[C_{0} +
C_{1}\frac{\sqrt{kT_{F1}}}{T_{D3}}T^{3}]\rightarrow \nonumber\\
&& T^{-1}\sim [ (\frac{40T_{F1}k\pi^{3}}{729})^{2/3}t^{2/3}+
(\frac{40T_{F1}k\pi^{3}}{729})^{2/11}t^{2/11}+ln(
\frac{16T_{D3}(T_{F1}k\pi)^{1/2}}{81}t)] ,\nonumber\\
&&\nonumber\\
&&V=\sigma_{0}=0,
\bar{C}_{0}=-C_{0}\rightarrow T_{end}=\frac{\bar{C}_{0}\sqrt{T_{D3}}}{C_{1}kT_{F1}}\rightarrow \nonumber\\
&&t_{end}\sim
\frac{C_{1}kT_{F1}}{\bar{C}_{0}\sqrt{T_{D3}}}[(\frac{40T_{F1}k\pi^{3}}{729})^{2/11}+\frac{16T_{D3}(T_{F1}k\pi)^{1/2}}{81}]^{-1}\label{m13}
\end{eqnarray}
This equation shows that temperature was infinity at the
beginning, decreased with time and tended to $T_{end}$ at large
values of time ($t=t_{end}$). Consequently, throat of a wormhole
is decreased with time and tends to zero at $t=t_{end}$.

After that, wormhole transfers energy from  extra
dimensions into our universe and causes  inflation.
Simultaneously, one stringy wormhole is formed in BIon. To
compare a stringy wormhole with a cylindrical wormhole, we will
construct a stringy wormhole in BIon.  Putting k units of F-string
charge along the radial direction and using equation (\ref{m7}),
we obtain \cite{q6,q7}:
\begin{eqnarray}
z(\sigma)= \int_{\sigma}^{\infty}
d\acute{\sigma}(\frac{F(\acute{\sigma})^{2}}{F(\sigma_{0})^{2}}-1)^{-\frac{1}{2}}
\label{m14}
\end{eqnarray}
In finite temperature BIon $F(\sigma)$ is given by
\begin{eqnarray}
F(\sigma) = \sigma^{2}\frac{4cosh^{2}\alpha - 3}{cosh^{4}\alpha}
\label{m15}
\end{eqnarray}
where $cosh\alpha$ is determined by the function:
\begin{eqnarray}
cosh^{2}\alpha = \frac{3}{2}\frac{cos\frac{\delta}{3} +
\sqrt{3}sin\frac{\delta}{3}}{cos\delta} \label{m16}
\end{eqnarray}
with the definitions:
\begin{eqnarray}
cos\delta \equiv \overline{T}^{4}\sqrt{1 +
\frac{k^{2}}{\sigma^{4}}},\, \overline{T} \equiv
(\frac{9\pi^{2}N}{4\sqrt{3}T_{D_{3}}})T, \, \kappa \equiv \frac{k
T_{F1}}{4\pi T_{D_{3}}} \label{m17}
\end{eqnarray}
In above equation, T is the finite temperature of BIon, N is the
number of D3-branes and $T_{D_{3}}$ and $T_{F1}$ are tensions of
brane and fundamental strings respectively. Attaching a mirror
solution to Eq. (\ref{m13}), we construct a wormhole configuration.
The separation distance $\bar{\Delta} = 2z(\sigma_{0})$ between
the N D3-branes and N anti D3-branes for a given brane-antibrane
wormhole configuration is defined by the four parameters N, k, T and
$\sigma_{0}$. We have:
\begin{eqnarray}
\bar{\Delta} = 2z(\sigma_{0})= 2\int_{\sigma_{0}}^{\infty}
d\acute{\sigma}(\frac{F(\acute{\sigma})^{2}}{F(\sigma_{0})^{2}}-1)^{-\frac{1}{2}}
\label{m18}
\end{eqnarray}
In the limit of small temperatures, we obtain:
\begin{eqnarray}
\bar{\Delta} =
\frac{2\sqrt{\pi}\Gamma(5/4)}{\Gamma(3/4)}\sigma_{0}(1 +
\frac{8}{27}\frac{k^{2}}{\sigma_{0}^{4}}\overline{T}^{8})
\label{m19}
\end{eqnarray}

  Let us now discuss the early inflationary model of universe in thermal BIon. For this, we
need to compute the contribution of the BIonic system with the four-
dimensional universe energy momentum tensor. The EM tensor for one
BIonic system with N D3-branes and k F-string charges is
\cite{q7},
 \begin{eqnarray}
&& T^{00}=\frac{2T_{D3}^{2}}{\pi
T^{4}}\frac{F(\sigma)}{\sqrt{F^{2}(\sigma)-F^{2}(\sigma_{0})}}\sigma^{2}\frac{4cosh^{2}\alpha
+ 1}{cosh^{4}\alpha} \nonumber \\&& T^{ii}=
-\gamma^{ii}\frac{8T_{D3}^{2}}{\pi
T^{4}}\frac{F(\sigma)}{\sqrt{F^{2}(\sigma)-F^{2}(\sigma_{0})}}\sigma^{2}\frac{1}{cosh^{2}\alpha},\,i=1,2,3
\nonumber \\&&T^{44}=\frac{2T_{D3}^{2}}{\pi
T^{4}}\frac{F(\sigma)}{F(\sigma_{0})}\sigma^{2}\frac{4cosh^{2}\alpha
+ 1}{cosh^{4}\alpha} \label{m20}
\end{eqnarray}
 This equation shows that with increasing temperature
in BIonic system, the energy-momentum tensors decreases. This is
due to the fact that when spikes of  branes and antibranes are well
separated, wormhole is not formed, so there isn't any channel for
flowing energy from universe branes into extra dimensions and
consequently the temperature is very high, however when two universe branes are
close to each other and connected by a wormhole, the temperature is
reduced to the lower values.

Also, in this model, we introduce two four dimensional  universes
that interact with each other via a wormhole and form a binary
system. In this model, z is the transverse direction
 between two universes and only the wormhole has a momentum in this
 direction.  To obtain the energy- momentum tensor in this system, we
use the Einstein's field equation in presence of fluid flow
that reads as:
\begin{equation}
{R_{ij}} - \frac{1}{2}{{\mathop{\rm g}\nolimits} _{ij}}R =
k{T_{ij}}. \label{m21}
\end{equation}
Using this equation, we can obtain the energy momentum tensor for
the universe-wormhole:
\begin{eqnarray}
&&  T^{00}= 6\frac{\dot{a}^{2}}{a^{2}}+
(\frac{D'}{D}+\frac{C'}{C})\sqrt{\Delta}\nonumber
\\&&T^{ii}= 4\frac{\ddot{a}}{a}
+2\frac{\dot{a}^{2}}{a^{2}}+\frac{1}{\sqrt{\Delta}}[2\ddot{V}+2\frac{B'}{B}\dot{V}^{2}+\frac{B'}{B^{2}}+\frac{D'}{D}\Delta],\,i=1,2,3
 \nonumber
\\&&T^{44}= \frac{1}{\sqrt{\Delta}}[2\ddot{V}+2\frac{B'}{B}\dot{V}^{2}+\frac{B'}{B^{2}}+\frac{C'}{C}\Delta]\label{m22}
\end{eqnarray}
On the other hand, such a higher-dimensional stress-energy tensor
is assumed to be that of a perfect fluid and of the form
 \begin{equation}
T_i^j = {\mathop{\rm diag}\nolimits} \left[ { - p, - p, - p, -
\bar{p}, - p, - p, - p, \rho } \right], \label{m23}
\end{equation}
 where $\bar{p}$ is the pressure in the extra space-like
 (z) dimension.  Using the energy momentum tensor of equations (\ref{m20}) and
(\ref{m22}) in the conservation law of equations (\ref{m21})  and
employing (\ref{m23}), we write:
\begin{eqnarray}
&&\rho
=\rho_{Uni1}+\rho_{Uni2}+\rho_{wormhole}=\rho_{BIon}\nonumber
\\&&
p_{i,tot}=p_{i,Uni1}+p_{i,Uni2}+p_{i,wormhole}=p_{i,BIon},\text{
   i=1,2,3} \nonumber
\\&& p_{z,tot}=p_{z,wormhole}=p_{z,BIon} \label{m24}
\end{eqnarray}
By using equations (\ref{m20}), (\ref{m22}) and (\ref{m24}), we
obtain:
\begin{eqnarray}
&&6\frac{\dot{a}^{2}}{a^{2}}+
(\frac{D'}{D}+\frac{C'}{C})\sqrt{\Delta}=\nonumber
\\&&\frac{2T_{D3}^{2}}{\pi
T^{4}}\frac{F(\sigma)}{\sqrt{F^{2}(\sigma)-F^{2}(\sigma_{0})}}\sigma^{2}\frac{4cosh^{2}\alpha
+ 1}{cosh^{4}\alpha}\nonumber
\\&&\nonumber
\\&&
4\frac{\ddot{a}}{a}
+2\frac{\dot{a}^{2}}{a^{2}}+\frac{1}{\sqrt{\Delta}}[2\ddot{V}+2\frac{B'}{B}\dot{V}^{2}+\frac{B'}{B^{2}}+\frac{D'}{D}\Delta]=\nonumber
\\&&-\frac{8T_{D3}^{2}}{\pi
T^{4}}\frac{F(\sigma)}{\sqrt{F^{2}(\sigma)-F^{2}(\sigma_{0})}}\sigma^{2}\frac{1}{cosh^{2}\alpha}
\nonumber
\\&& \nonumber
\\&&\frac{1}{\sqrt{\Delta}}[2\ddot{V}+2\frac{B'}{B}\dot{V}^{2}+\frac{B'}{B^{2}}+\frac{C'}{C}\Delta]=\nonumber
\\&&-\frac{2T_{D3}^{2}}{\pi
T^{4}}\frac{F(\sigma)}{F(\sigma_{0})}\sigma^{2}\frac{4cosh^{2}\alpha
+ 1}{cosh^{4}\alpha}  \label{m25}
\end{eqnarray}
Assuming D=C and with the help of equation (\ref{m12}), we solve the
above equation and obtain the explicit forms of a,B,C and D :
\begin{eqnarray}
&&a(t)=exp(\int U dt)\nonumber
\\&& U=-\frac{8T_{D3}^{2}}{\pi
T^{4}}\frac{F(\sigma)}{\sqrt{F^{2}(\sigma)-F^{2}(\sigma_{0})}}\sigma^{2}\frac{1}{cosh^{2}\alpha}+
\frac{2T_{D3}^{2}}{\pi
T^{4}}\frac{F(\sigma)}{F(\sigma_{0})}\sigma^{2}\frac{4cosh^{2}\alpha
+ 1}{cosh^{4}\alpha}\rightarrow \nonumber
\\&& U= -\frac{8T_{D3}^{2}}{\pi
T^{4}}\frac{\sigma^{4}}{\sqrt{\sigma^{2}-\sigma_{0}^{2}}}(\frac{2T^{4}\sqrt{1+\frac{\beta^{2}}{\sigma^{4}}}}{3\sqrt{3}-T^{4}\sqrt{1+\frac{\beta^{2}}{\sigma^{4}}}
-\frac{\sqrt{3}}{6}T^{8}(1+\frac{\beta^{2}}{\sigma^{4}})})^{2}
+\nonumber
\\&&
\frac{2T_{D3}^{2}}{\pi
T^{4}}\frac{\sigma^{4}}{\sigma_{0}^{2}}(4+(\frac{2T^{4}\sqrt{1+\frac{\beta^{2}}{\sigma^{4}}}}{3\sqrt{3}-T^{4}\sqrt{1+\frac{\beta^{2}}{\sigma^{4}}}
-\frac{\sqrt{3}}{6}T^{8}(1+\frac{\beta^{2}}{\sigma^{4}})})^{4})\rightarrow
\nonumber
\\&& U= -[\frac{8T_{D3}^{2}[(\frac{40T_{F1}k\pi^{3}}{729})^{2/3}t^{2/3}+
(\frac{40T_{F1}k\pi^{3}}{729})^{2/11}t^{2/11}+ln(
\frac{16T_{D3}(T_{F1}k\pi)^{1/2}}{81}t)]^{4}}{\pi
}\frac{\sigma^{4}}{\sqrt{\sigma^{2}-\sigma_{0}^{2}}}\times
\nonumber
\\&&(\frac{2[
(\frac{40T_{F1}k\pi^{3}}{729})^{2/3}t^{2/3}+
(\frac{40T_{F1}k\pi^{3}}{729})^{2/11}t^{2/11}+ln(
\frac{16T_{D3}(T_{F1}k\pi)^{1/2}}{81}t)]^{-4}\sqrt{1+\frac{\beta^{2}}{\sigma^{4}}}}{3\sqrt{3}-t^{-8/3}\sqrt{1+\frac{\beta^{2}}{\sigma^{4}}}
-\frac{\sqrt{3}}{6}[ (\frac{40T_{F1}k\pi^{3}}{729})^{2/3}t^{2/3}+
(\frac{40T_{F1}k\pi^{3}}{729})^{2/11}t^{2/11}+ln(
\frac{16T_{D3}(T_{F1}k\pi)^{1/2}}{81}t)]^{-8}(1+\frac{\beta^{2}}{\sigma^{4}})})^{2}]
\nonumber
\\&&
+[\frac{2T_{D3}^{2}[(\frac{40T_{F1}k\pi^{3}}{729})^{2/3}t^{2/3}+
(\frac{40T_{F1}k\pi^{3}}{729})^{2/11}t^{2/11}+ln(
\frac{16T_{D3}(T_{F1}k\pi)^{1/2}}{81}t)]^{4}}{\pi
}\frac{\sigma^{4}}{\sigma_{0}^{2}}(4+ \nonumber
\\&&(\frac{2(\frac{40T_{F1}k\pi^{3}}{729})^{2/3}t^{2/3}+
(\frac{40T_{F1}k\pi^{3}}{729})^{2/11}t^{2/11}+ln(
\frac{16T_{D3}(T_{F1}k\pi)^{1/2}}{81}t)]^{-4}\sqrt{1+\frac{\beta^{2}}{\sigma^{4}}}}{3\sqrt{3}-[t]^{-8/3}\sqrt{1+\frac{\beta^{2}}{\sigma^{4}}}
-\frac{\sqrt{3}}{6}(\frac{40T_{F1}k\pi^{3}}{729})^{2/3}t^{2/3}+
(\frac{40T_{F1}k\pi^{3}}{729})^{2/11}t^{2/11}+ln(
\frac{16T_{D3}(T_{F1}k\pi)^{1/2}}{81}t)]^{-8}(1+\frac{\beta^{2}}{\sigma^{4}})})^{4})]\nonumber
\\&&\nonumber
\\&&
D=C=exp(\int E dr)\nonumber
\\&& E=-\frac{2T_{D3}^{2}}{\pi
T^{4}}\frac{F(\sigma)}{F(\sigma_{0})}\sigma^{2}\frac{4cosh^{2}\alpha
+ 1}{cosh^{4}\alpha}+\frac{46T_{D3}^{2}}{\pi
T^{4}}\frac{F(\sigma)}{\sqrt{F^{2}(\sigma)-F^{2}(\sigma_{0})}}\sigma^{2}\frac{1}{cosh^{2}\alpha}\rightarrow\nonumber
\\&&E= -[\frac{2T_{D3}^{2}[(\frac{40T_{F1}k\pi^{3}}{729})^{2/3}t^{2/3}+
(\frac{40T_{F1}k\pi^{3}}{729})^{2/11}t^{2/11}+ln(
\frac{16T_{D3}(T_{F1}k\pi)^{1/2}}{81}t)]^{4}}{\pi
}\frac{\sigma^{4}}{\sigma_{0}^{2}}(4+ \nonumber
\\&&(\frac{2(\frac{40T_{F1}k\pi^{3}}{729})^{2/3}t^{2/3}+
(\frac{40T_{F1}k\pi^{3}}{729})^{2/11}t^{2/11}+ln(
\frac{16T_{D3}(T_{F1}k\pi)^{1/2}}{81}t)]^{-4}\sqrt{1+\frac{\beta^{2}}{\sigma^{4}}}}{3\sqrt{3}-[t]^{-8/3}\sqrt{1+\frac{\beta^{2}}{\sigma^{4}}}
-\frac{\sqrt{3}}{6}(\frac{40T_{F1}k\pi^{3}}{729})^{2/3}t^{2/3}+
(\frac{40T_{F1}k\pi^{3}}{729})^{2/11}t^{2/11}+ln(
\frac{16T_{D3}(T_{F1}k\pi)^{1/2}}{81}t)]^{-8}(1+\frac{\beta^{2}}{\sigma^{4}})})^{4})]\nonumber
\\&&+[\frac{48T_{D3}^{2}[(\frac{40T_{F1}k\pi^{3}}{729})^{2/3}t^{2/3}+
(\frac{40T_{F1}k\pi^{3}}{729})^{2/11}t^{2/11}+ln(
\frac{16T_{D3}(T_{F1}k\pi)^{1/2}}{81}t)]^{4}}{\pi
}\frac{\sigma^{4}}{\sqrt{\sigma^{2}-\sigma_{0}^{2}}}\times
\nonumber
\\&&(\frac{2[
(\frac{40T_{F1}k\pi^{3}}{729})^{2/3}t^{2/3}+
(\frac{40T_{F1}k\pi^{3}}{729})^{2/11}t^{2/11}+ln(
\frac{16T_{D3}(T_{F1}k\pi)^{1/2}}{81}t)]^{-4}\sqrt{1+\frac{\beta^{2}}{\sigma^{4}}}}{3\sqrt{3}-t^{-8/3}\sqrt{1+\frac{\beta^{2}}{\sigma^{4}}}
-\frac{\sqrt{3}}{6}[ (\frac{40T_{F1}k\pi^{3}}{729})^{2/3}t^{2/3}+
(\frac{40T_{F1}k\pi^{3}}{729})^{2/11}t^{2/11}+ln(
\frac{16T_{D3}(T_{F1}k\pi)^{1/2}}{81}t)]^{-8}(1+\frac{\beta^{2}}{\sigma^{4}})})^{2}]\nonumber
\\&&\rightarrow E\simeq-U \nonumber
\\&& \nonumber
\\&&B=exp(\int w dr)\nonumber
\\&&w=-(\frac{1}{1+2\dot{P}^{2}}) (\frac{2T_{D3}^{2}}{\pi
T^{4}}\frac{F(\sigma)}{F(\sigma_{0})}\sigma^{2}\frac{4cosh^{2}\alpha
+ 1}{cosh^{4}\alpha})-2\ddot{U}\rightarrow\nonumber
\\&&w=-2\ddot{U}-[\frac{2T_{D3}^{2}[(\frac{40T_{F1}k\pi^{3}}{729})^{2/3}t^{2/3}+
(\frac{40T_{F1}k\pi^{3}}{729})^{2/11}t^{2/11}+ln(
\frac{16T_{D3}(T_{F1}k\pi)^{1/2}}{81}t)]^{4}}{\pi
}\frac{\sigma^{4}}{\sigma_{0}^{2}}(4+ \nonumber
\\&&(\frac{2(\frac{40T_{F1}k\pi^{3}}{729})^{2/3}t^{2/3}+
(\frac{40T_{F1}k\pi^{3}}{729})^{2/11}t^{2/11}+ln(
\frac{16T_{D3}(T_{F1}k\pi)^{1/2}}{81}t)]^{-4}\sqrt{1+\frac{\beta^{2}}{\sigma^{4}}}}{3\sqrt{3}-[t]^{-8/3}\sqrt{1+\frac{\beta^{2}}{\sigma^{4}}}
-\frac{\sqrt{3}}{6}(\frac{40T_{F1}k\pi^{3}}{729})^{2/3}t^{2/3}+
(\frac{40T_{F1}k\pi^{3}}{729})^{2/11}t^{2/11}+ln(
\frac{16T_{D3}(T_{F1}k\pi)^{1/2}}{81}t)]^{-8}(1+\frac{\beta^{2}}{\sigma^{4}})})^{4})]\nonumber
\\&&\rightarrow
w\simeq -U \label{m26}
\end{eqnarray}
This equation indicates that while the temperature is decreased, the
parameters of wormhole are reduced to lower values and tended to zero
at $\sigma_{0}=V=0$ and $t=t_{end}$(see Figures 1 and 2). This
means that the wormhole is disappeared at the end of inflation, however
the scale factor of universe is increased very fast and tended to
large values in this epoch(see figures 3.).

\begin{figure}
\centering
\includegraphics[height=6cm]{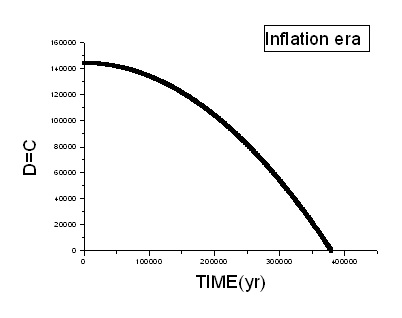}
\caption{The wormhole parameter (D=C)  for inflation era of
expansion history as a function of t where t is the age of universe.
In this plot, we choose $t_{birth}=0$, $t_{end}$=380000,
$T_{birth}=10^{32}$ and $T_{end}=10^{^{9}}$.} \label{fig:figure1}
\end{figure}
\begin{figure}
\centering
\includegraphics[height=6cm]{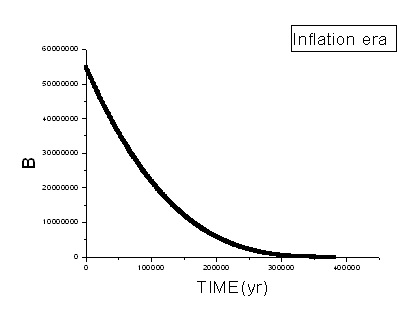}
\caption{The wormhole parameter (B)  for inflation era of
expansion history as a function of t where t is age of universe.
In this plot, we choose $t_{birth}=0$, $t_{end}$=380000,
$T_{birth}=10^{32}$ and $T_{end}=10^{^{9}}$.} \label{fig:figure2}
\end{figure}
\begin{figure}
\centering
\includegraphics[height=6cm]{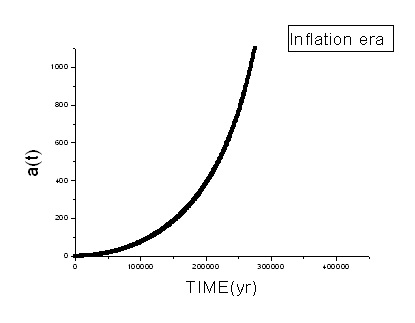}
\caption{The scale factor (a)  for inflation era of expansion
history as a function of t where t is age of universe. In this
plot, we choose $t_{birth}=0$, $t_{end}$=380000,
$T_{birth}=10^{32}$ and $T_{end}=10^{^{9}}$.} \label{fig:figure2}
\end{figure}

\section{The birth and death of cylindrical wormhole during late time acceleration }\label{o2}
In this section, we discuss that with decreasing the separation
distance between brane-antibrane, a tachyon is born, grows very fast
and causes formation of a new cylindrical wormhole. This
wormhole transfers energy from extra dimensions into our
universe according to which the second phase of acceleration takes place.

To construct a non-phantom model, we consider a set of
D3-$\overline{D3}$-brane pairs in the background  (\ref{m7}) which
are placed at points $z_{1} = l/2$ and $z_{2} = -l/2$ respectively
so that the separation between the brane and antibrane is l. For
the simple case of a single D3-$\overline{D3}$-brane pair with an
open string tachyon, the action is\cite{q15,q16}:
 \begin{eqnarray}
&& S=-\tau_{3}\int d^{9}\sigma \sum_{i=1}^{2}
V(TA,l)e^{-\phi}(\sqrt{-det A_{i}})\nonumber \\&&
(A_{i})_{ab}=(g_{MN}-\frac{TA^{2}l^{2}}{Q}g_{Mz}g_{zN})\partial_{a}x^{M}_{i}\partial_{b}x^{M}_{i}
+F^{i}_{ab}+\frac{1}{2Q}((D_{a}TA)(D_{b}TA)^{\ast}+(D_{a}TA)^{\ast}(D_{b}TA))\nonumber
\\&&
+il(g_{az}+\partial_{a}z_{i}g_{zz})(TA(D_{b}TA)^{\ast}-TA^{\ast}(D_{b}TA))+
il(TA(D_{a}TA)^{\ast}-TA^{\ast}(D_{a}TA))\times\nonumber
\\&&(g_{bz}+\partial_{b}z_{i}g_{zz})(1-\frac{\pi^{2}N
T^{4}}{6T_{D3}}). \label{m27}
\end{eqnarray}
where
  \begin{eqnarray}
&& Q=1+TA^{2}l^{2}g_{zz}, \nonumber \\&&
D_{a}TA=\partial_{a}TA-i(A_{2,a}-A_{1,a})TA,
V(TA,l)=g_{s}V(TA)\sqrt{Q}, \nonumber \\&& e^{\phi}=g_{s}( 1 +
\frac{R^{4}}{z^{4}} )^{-\frac{1}{2}}, \label{m28}
\end{eqnarray}
$\phi$, $A_{2,a}$ and $F^{i}_{ab}$ are dilaton field, the gauge
field and field strength on the world-volume of the non-BPS
brane respectively, TA is the tachyon field, $\tau_{3}$ is the brane tension
and V (TA) is the tachyon potential. The indices a,b denote the
tangent directions of D-branes, while the indices M,N run over the
background of ten-dimensional space-time directions. The Dp-brane and
the anti-Dp-brane are labeled by i = 1 and 2 respectively. Then
the separation between these D-branes is defined by $z_{2} - z_{1}
= l$. Also, in writing the above equations we are using the convention
$2\pi\acute{\alpha}=1$. A potential which has been used in most
papers is \cite{q17,q18,q19}:
 \begin{eqnarray}
V(TA)=\frac{\tau_{3}}{cosh\sqrt{\pi}TA} \label{m29}
\end{eqnarray}

 Let us consider the only $\sigma$
dependence of the tachyon field TA for simplicity and set the
gauge fields to zero. In this case, the action (\ref{m27}) in the
region  $r> R$ and $TA'\sim constant$ simplifies to
   \begin{eqnarray}
L \simeq-\frac{\tau_{3}}{g_{s}} \int d\sigma \sigma^{2}
V(TA)(\sqrt{D_{1,TA}}+\sqrt{D_{2,TA}})(1-\frac{\pi^{2}N
T^{4}}{6T_{D3}}) \label{m30}
\end{eqnarray}
where
 \begin{eqnarray}
D_{1,TA} = D_{2,TA}\equiv D_{TA} = 1 + \frac{l'(\sigma)^{2}}{4}+
\dot{TA}^{2} -  TA'^{2} \label{m31}
\end{eqnarray}
 we assume that $TA l\ll TA'$. Now, we  study the Hamiltonian
corresponding to the above Lagrangian.
 To derive this we need the canonical momentum density $\Pi =
\frac{\partial L}{\partial \dot{TA}}$ associated with the tachyon:
 \begin{eqnarray}
\Pi = \frac{V(TA)\dot{TA}}{ \sqrt{1 + \frac{l'(\sigma)^{2}}{4}+
\dot{TA}^{2} -  TA'^{2}}}(1-\frac{\pi^{2}N
T^{4}}{6T_{D3}})\label{m32}
\end{eqnarray}
so that the Hamiltonian can be obtained as:
\begin{eqnarray}
H_{DBI} = 4\pi\int d\sigma  \sigma^{2} \Pi \dot{TA} - L
 \label{m33}
\end{eqnarray}
By choosing $\dot{TA} = 2 TA'$, this gives:
\begin{eqnarray}
H_{DBI} = 4\pi\int d\sigma \sigma^{2} [\Pi
(\dot{TA}-\frac{1}{2}TA')] + \frac{1}{2}TA\partial_{\sigma}(\Pi
\sigma^{2}) - L
 \label{m34}
\end{eqnarray}
In this equation, we have in the second step integration by parts
the term proportional to $\dot{TA}$, indicating that a tachyon can
be studied as a Lagrange multiplier imposing the constraint
$\partial_{\sigma}(\Pi \sigma^{2}V(TA))=0$ on the canonical
momentum. Solving this equation yields:
\begin{eqnarray}
\Pi =\frac{\beta}{4\pi \sigma^{2}}
 \label{m35}
\end{eqnarray}
where $\beta$ is a constant.  Using (\ref{m35}) in (\ref{m34}), we
get:
\begin{eqnarray}
&& H_{DBI} = \int d\sigma V(TA)\sqrt{1 + \frac{l'(\sigma)^{2}}{4}
+ \dot{TA}^{2} -  TA'^{2}}F_{DBI} ,  \nonumber \\&&
F_{DBI}=\sigma^{2}\sqrt{1 +
\frac{\beta}{\sigma^{2}}}(1-\frac{\pi^{2}N
T^{4}}{6T_{D3}})\label{m36}
\end{eqnarray}
The output of EOM for $l(\sigma)$, calculated by varying
(\ref{m36}), is
\begin{eqnarray}
&&(\frac{l'F_{DBI}}{4\sqrt{1+
\frac{l'(\sigma)^{2}}{4}}})'=0\label{m37}
\end{eqnarray}
Solving this equation, we obtain:
\begin{eqnarray}
&&l(\sigma) = 2(\frac{l_{0}}{2} -\int_{\sigma}^{\infty} d\sigma
(\frac{F_{DBI}(\sigma)}{F_{DBI}(\sigma_{0})}-1)^{-\frac{1}{2}})
\label{m38}
\end{eqnarray}
This solution, for non-zero $\sigma_{0}$  represents a wormhole
with a finite size throat. This equation indicates that the
separation distance between two branes is $l_{0}$ at the birth of
wormhole($\sigma_{0}$), decreases with time and shrinks to zero at
larger values of throat. On the hand, to obtain the explicit form
of a tachyon , we are using the equation of motion extracted from
action (\ref{m30}):
\begin{eqnarray}
(\frac{1}{\sqrt{D_{TA}}}TA'(\sigma)\acute{)}=\frac{1}{\sqrt{D_{TA}}}
[\frac{(V(TA)F_{DBI})}{F_{DBI}V(TA)'}(D_{TA}-TA'(\sigma)^{2})]
\label{m39}
\end{eqnarray}
with a solution
\begin{eqnarray}
TA\sim
\sqrt{\frac{\sigma_{0}^{2}}{\sigma_{0}^{2}-\sigma^{2}}}(\frac{1}{1+\frac{\pi^{2}N
T^{4}}{6T_{D3}}} )\label{m40}
\end{eqnarray}
This equation shows that a tachyon is zero before the birth of
a wormhole( $\sigma_{0}=0$) and with a decrease in temperature and
an increase in the throat of a wormhole it grows to larger
values.

At this stage, we consider the late time acceleration of universe in
thermal BIon. To achieve this, we calculate the contribution of a tachyonic
wormhole to the four- dimensional universe energy momentum tensor.
We have:

 \begin{eqnarray}
&& T^{00}=V(TA)F_{DBI}\sqrt{D_{TA}},  \nonumber \\&&
T^{44}=-V(TA)F_{DBI}\frac{1}{\sqrt{D_{TA}}}
(TA^{2}l^{2}+\frac{\acute{l}^{2}}{4}) \nonumber \\&& T^{ii} =
-V(TA)F_{DBI}\frac{Q}{\sqrt{D_{TA}}},\,i=1,2,3 \label{m41}
\end{eqnarray}
Setting the energy momentum tensor of equations (\ref{m22}) and
(\ref{m41}) in the conservation law of equations (\ref{m21}) and
(\ref{m24}) and employing (\ref{m23}) yields:
\begin{eqnarray}
&&6\frac{\dot{a}^{2}}{a^{2}}+
(\frac{D'}{D}+\frac{C'}{C})\sqrt{\Delta}=\nonumber
\\&&V(TA)F_{DBI}\sqrt{D_{TA}}\nonumber
\\&&\nonumber
\\&&
4\frac{\ddot{a}}{a}
+2\frac{\dot{a}^{2}}{a^{2}}+\frac{1}{\sqrt{\Delta}}[2\ddot{V}+2\frac{B'}{B}\dot{V}^{2}+\frac{B'}{B^{2}}+\frac{D'}{D}\Delta]=\nonumber
\\&&-V(TA)F_{DBI}\frac{Q}{\sqrt{D_{TA}}}\nonumber
\\&& \nonumber
\\&&\frac{1}{\sqrt{\Delta}}[2\ddot{V}+2\frac{B'}{B}\dot{V}^{2}+\frac{B'}{B^{2}}+\frac{C'}{C}\Delta]=\nonumber
\\&&-V(TA)F_{DBI}\frac{1}{\sqrt{D_{TA}}}
(TA^{2}l^{2}+\frac{\acute{l}^{2}}{4})  \label{m42}
\end{eqnarray}
Solving these equations simultaneously, assuming (D=C),
$V(t)=exp(-\int P_{0}dt)$ and using equations (\ref{m12}),
(\ref{m38}) and (\ref{m40}), we obtain the explicit form of
wormhole parameters and the scale factor of universe:
\begin{eqnarray}
&&a(t)=exp(\int dt G(t))\nonumber
\\&&G(t)=exp(\frac{6T_{D3}
[ (\frac{40T_{F1}k\pi^{3}}{729})^{2/3}t^{2/3}+
(\frac{40T_{F1}k\pi^{3}}{729})^{2/11}t^{2/11}+ln(
\frac{16T_{D3}(T_{F1}k\pi)^{1/2}}{81}t)]^{4}}{\pi^{2}N})\times\nonumber
\\&&(\frac{6T_{D3}
[ (\frac{40T_{F1}k\pi^{3}}{729})^{2/3}t^{2/3}+
(\frac{40T_{F1}k\pi^{3}}{729})^{2/11}t^{2/11}+ln(
\frac{16T_{D3}(T_{F1}k\pi)^{1/2}}{81}t)]^{4}}{\pi^{2}N})\sqrt{\frac{\sigma_{0}^{2}}{\sigma_{0}^{2}-\sigma^{2}}}
\nonumber
\\&&\nonumber
\\&&
D=C= exp(\int H dr)\nonumber
\\&& H= \sqrt{1+(\frac{\sigma_{0}^{2}}{\sigma_{0}^{2}-\sigma^{2}})(1+\frac{\pi^{2}N
[ (\frac{40T_{F1}k\pi^{3}}{729})^{2/3}t^{2/3}+
(\frac{40T_{F1}k\pi^{3}}{729})^{2/11}t^{2/11}+ln(
\frac{16T_{D3}(T_{F1}k\pi)^{1/2}}{81}t)]^{4}}{6T_{D3}}})^{2}\times\nonumber
\\&&exp(\frac{6T_{D3}
[ (\frac{40T_{F1}k\pi^{3}}{729})^{2/3}t^{2/3}+
(\frac{40T_{F1}k\pi^{3}}{729})^{2/11}t^{2/11}+ln(
\frac{16T_{D3}(T_{F1}k\pi)^{1/2}}{81}t)]^{4}}{\pi^{2}N})\times\nonumber
\\&&(\frac{6T_{D3}
[ (\frac{40T_{F1}k\pi^{3}}{729})^{2/3}t^{2/3}+
(\frac{40T_{F1}k\pi^{3}}{729})^{2/11}t^{2/11}+ln(
\frac{16T_{D3}(T_{F1}k\pi)^{1/2}}{81}t)]^{4}}{\pi^{2}N})\sigma^{2}\sqrt{1
+ \frac{\beta}{\sigma^{2}}}\nonumber
\\&&\nonumber
\\&&B= exp(\int X dr)\nonumber
\\&& X=(\frac{[ (\frac{40T_{F1}k\pi^{3}}{729})^{2/3}t^{2/3}+
(\frac{40T_{F1}k\pi^{3}}{729})^{2/11}t^{2/11}+ln(
\frac{16T_{D3}(T_{F1}k\pi)^{1/2}}{81}t)]^{2}4(\frac{l_{0}}{2}
-\int_{\sigma}^{\infty} d\sigma
(\frac{\sigma^{2}}{\sigma_{0}^{2}}-1)^{-1/2})^{2}
}{\sqrt{(\frac{\sigma_{0}^{2}}{\sigma_{0}^{2}-\sigma^{2}})+[
(\frac{40T_{F1}k\pi^{3}}{729})^{2/3}t^{2/3}+
(\frac{40T_{F1}k\pi^{3}}{729})^{2/11}t^{2/11}+ln(
\frac{16T_{D3}(T_{F1}k\pi)^{1/2}}{81}t)]^{2}}}\nonumber
\\&&+\frac{\frac{\sigma_{0}^{2}}{\sigma_{0}^{2}-\sigma^{2}}}{\sqrt{(\frac{\sigma_{0}^{2}}{\sigma_{0}^{2}-\sigma^{2}})+[
(\frac{40T_{F1}k\pi^{3}}{729})^{2/3}t^{2/3}+
(\frac{40T_{F1}k\pi^{3}}{729})^{2/11}t^{2/11}+ln(
\frac{16T_{D3}(T_{F1}k\pi)^{1/2}}{81}t)]^{2}}})\times\nonumber
\\&&exp(\frac{6T_{D3}
[ (\frac{40T_{F1}k\pi^{3}}{729})^{2/3}t^{2/3}+
(\frac{40T_{F1}k\pi^{3}}{729})^{2/11}t^{2/11}+ln(
\frac{16T_{D3}(T_{F1}k\pi)^{1/2}}{81}t)]^{4}}{\pi^{2}N})\times\nonumber
\\&&(\frac{6T_{D3}
[ (\frac{40T_{F1}k\pi^{3}}{729})^{2/3}t^{2/3}+
(\frac{40T_{F1}k\pi^{3}}{729})^{2/11}t^{2/11}+ln(
\frac{16T_{D3}(T_{F1}k\pi)^{1/2}}{81}t)]^{4}}{\pi^{2}N})\sigma^{2}\sqrt{1
+ \frac{\beta}{\sigma^{2}}}
 \label{m43}
\end{eqnarray}
These solutions indicate that the wormhole parameters and also the scale factor
are increased with time and shrink to infinity at
($\sigma=\sigma_{0}$, $t=t_{rip}$). In figures 1,2 and 3, we
compare wormhole parameters and the scale factor. As can be seen from
these figures, the value of scale factor is higher than the corresponding values for the wormhole
parameters at the beginning of acceleration era. On the other hand, the rate
of growth for the wormhole parameters is more considerable. For this reason, we
predict that these parameters overtake the scale factor and lead
to the destruction of universe at future big rip singularity
\cite{q20}.

\begin{figure}
\centering
\includegraphics[height=6cm]{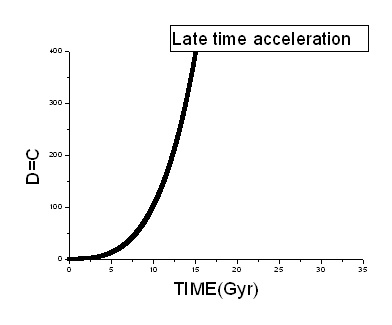}
\caption{The wormhole parameter (D=C)  for late time acceleration
era of expansion history as a function of t where t is age of
universe. In this plot, we choose $t_{late}=.4 Gyr$, $t_{rip}$=33
Gyr, $T_{late}=10^{4}$ and $T_{end}=0$.} \label{fig:figure1}
\end{figure}
\begin{figure}
\centering
\includegraphics[height=6cm]{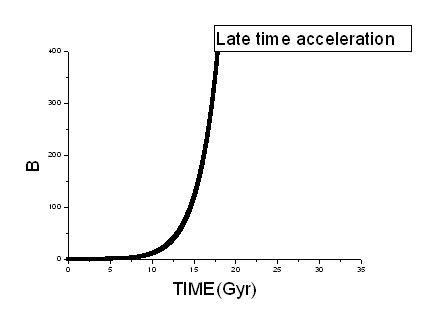}
\caption{The wormhole parameter (B)  for late time acceleration
era of expansion history as a function of t where t is age of
universe. In this plot, we choose $t_{late}=.4 Gyr$, $t_{rip}$=33
Gyr, $T_{late}=10^{4}$ and $T_{end}=0$.} \label{fig:figure2}
\end{figure}
\begin{figure}
\centering
\includegraphics[height=6cm]{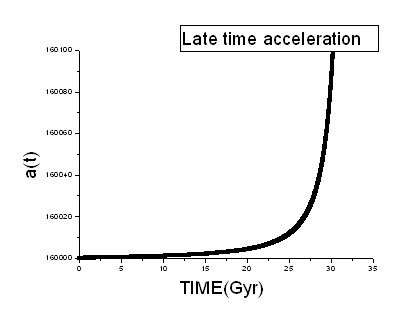}
\caption{The scale factor (a)  for late time acceleration era of
expansion history as a function of t where t is age of universe.
In this plot, we choose $t_{late}=.4 Gyr$, $t_{rip}$=33 Gyr,
$T_{late}=10^{4}$ and $T_{end}=0$.} \label{fig:figure2}
\end{figure}

\section{Summary and Discussion} \label{sum}
Recently, the stability analysis of cylindrical thin shell
wormholes has been
   studied in the literature.
  In this paper, we have proposed a new model that allows us to account for dynamics of this
  wormhole during different epochs of cosmic history from inflation
to recent observed acceleration era. In this model, coincided with
the birth of universe at the corresponding point, the early wormhole
is born. At this point, $k$ black fundamental strings are transited to
BIon which is a configuration of a universe brane and a universe
anti-brane connected by a wormhole. This wormhole transfers
energy from another universe to our own universe and causes
inflation. We have shown that two universe-branes can be connected by
an unstable cylindrical thin shell wormhole that vanishes very
fast. After the wormhole death, there isn't any channel for
flowing energy into our universe brane, inflation ends and a non
phantom era begins. With decreasing the separation between universe
branes, the second type of cylindrical thin shell wormholes, named as tachyonic wormholes are created. In this condition,
two universe branes are connected again and late time acceleration era
is started. After that we have considered the stability of these wormholes and
came to a conclusion that they will vanish at a future singularity.

\section*{Acknowledgments}
\noindent Authors wish to thank Professor S. Habib Mazharimousavi
and Professor M. Halilsoy for their nice comments that help us to
improve our paper.

 \end{document}